\documentclass[aps,prl,twocolumn,preprintnumbers,10pt,showpacs,nofootinbib]{revtex4-1}

\usepackage{amsmath,amsfonts,epsfig,color,latexsym}
\usepackage{amssymb}
\usepackage[english, USenglish]{babel}
\usepackage{epsfig,psfrag}
\usepackage{slashed}
\usepackage[normalem]{ulem}
\usepackage{color}

\usepackage{soul,xcolor}
\usepackage{hyperref}
\setstcolor{red}
\usepackage{comment}

\begin{document}
\renewcommand{\thefootnote}{\fnsymbol{footnote}}
\newpage
\pagestyle{empty}
\setcounter{page}{0}


\newcommand{\norm}[1]{{\protect\normalsize{#1}}}
\newcommand{\p}[1]{(\ref{#1})}
\newcommand{\half}{\tfrac{1}{2}}
\newcommand \vev [1] {\langle{#1}\rangle}
\newcommand \ket [1] {|{#1}\rangle}
\newcommand \bra [1] {\langle {#1}|}
\newcommand \pd [1] {\frac{\pa}{\pa {#1}}}
\newcommand \ppd [2] {\frac{\pa^2}{\pa {#1} \pa{#2}}}
\newcommand{\ed}[1]{{\color{red} {#1}}}

\newcommand{\cI}{{\cal I}}
\newcommand{\cM}{{\cal M}} 
\newcommand{\cR}{{\cal R}} 
\newcommand{\cS}{{\cal S}} 
\newcommand{\cK}{{\cal K}}
\newcommand{\cL}{{\cal L}} 
\newcommand{\cF}{{\cal F}}
\newcommand{\cN}{{\cal N}}
\newcommand{\cA}{{\cal A}}
\newcommand{\cB}{{\cal B}}
\newcommand{\cG}{{\cal G}}
\newcommand{\cO}{{\cal O}}
\newcommand{\cY}{{\cal Y}}
\newcommand{\cX}{{\cal X}}
\newcommand{\cT}{{\cal T}}
\newcommand{\cW}{{\cal W}}
\newcommand{\cP}{{\cal P}}
\newcommand{\bP}{{\bar\Phi}}
\newcommand{\mK}{{\mathbb K}}
\newcommand{\nt}{\notag\\} 
\newcommand{\pa}{\partial}
\newcommand{\ep}{\epsilon}
\newcommand{\om}{\omega}
\newcommand{\bom}{\bar\omega}
\newcommand{\etap}{\bar\epsilon}
\newcommand{\vep}{\varepsilon}
\renewcommand{\a}{\alpha}
\renewcommand{\b}{\beta}
\newcommand{\g}{\gamma}
\newcommand{\s}{\sigma}
\newcommand{\la}{\lambda}
\newcommand{\tl}{\tilde\lambda}
\newcommand{\tm}{\tilde\mu}
\newcommand{\tk}{\tilde k}
\newcommand{\da}{{\dot\alpha}}
\newcommand{\db}{{\dot\beta}}
\newcommand{\dg}{{\dot\gamma}}
\newcommand{\dd}{{\dot\delta}}
\newcommand{\q}{\theta}
\newcommand{\bq}{{\bar\theta}}
\renewcommand{\r}{\rho}
\newcommand{\br}{\bar\rho}
\newcommand{\be}{\bar\eta}
\newcommand{\bQ}{\bar Q}
\newcommand{\bx}{\bar \xi}
\newcommand{\tx}{\tilde{x}}
\newcommand{\tr}{\mbox{tr}}
\newcommand{\+}{{\dt+}}
\renewcommand{\-}{{\dt-}}
\newcommand{\ti}{{\textup{i}}}

\newcommand{\dlog}{d{\rm log}}
\newcommand{\tred}[1]{\textcolor{red}{\bfseries #1}}
\newcommand{\eps}{\epsilon}

\preprint{MPP-2018-305, ZU-TH 49/18, MITP/19-003}

\title{
Analytic result for a two-loop five-particle amplitude 
}

\author{D.\ Chicherin$^{a}$, T.\ Gehrmann$^{b}$, J.\ M.\ Henn$^{a}$, P.\ Wasser$^{c}$, Y.\ Zhang$^{a}$, S.\ Zoia$^{a}$}

\affiliation{
$^a$ Max-Planck-Institut f{\"u}r Physik, Werner-Heisenberg-Institut, D-80805 M{\"u}nchen, Germany\\
$^b$ Physik-Institut, Universit{\"a}t Z{\"u}rich, Wintherturerstrasse 190, CH-8057 Z{\"u}rich, Switzerland\\
$^c$ PRISMA Cluster of Excellence, 
Johannes Gutenberg University, D-55099 Mainz, Germany
}

\begin{abstract}
We compute the symbol of the full-color two-loop five-particle amplitude in $\mathcal{N}=4$ super Yang-Mills, including all non-planar subleading-color terms.
The amplitude is written in terms of permutations of Parke-Taylor tree-level amplitudes and pure functions to all orders in the dimensional regularization parameter, in agreement with previous conjectures.
The answer has the correct collinear limits and infrared factorization properties, allowing us to define a finite remainder function.
We study the multi-Regge limit of the non-planar terms, analyze its subleading power corrections, and present analytically the leading logarithmic terms.

\end{abstract}

\maketitle

The study of scattering amplitudes in maximally supersymmetric Yang-Mills theory ($\cN = 4$ sYM) has brought about many advances in quantum field theory (QFT). Experience shows that having analytical `data', i.e. explicit results, for amplitudes available is vital to find structures and patterns in seemingly complicated results, and to test new ideas. Cases in point are dual-conformal symmetry~\cite{Drummond:2008vq,Drummond:2009fd,Berkovits:2008ic}, the symbol analysis~\cite{Goncharov:2010jf}, insights of Regge limits in perturbative QFT~\cite{Bartels:2008ce}, and the structure of infrared divergences~\cite{vanNeerven:1985ja,Bern:2005iz}, just to name a few.  

Thanks to recent progress, an abundant wealth of data is available for planar scattering amplitudes in $\cN = 4$ sYM. Up to five particles, the functional form of the latter is fixed by dual conformal symmetry~\cite{Drummond:2006rz,Drummond:2007au}, in agreement with previous conjectures~\cite{Anastasiou:2003kj,Bern:2005iz}. Starting from six particles, there is a freedom of a dual conformally invariant function~\cite{Drummond:2008vq,Drummond:2008aq,Bern:2008ap}, which has been the subject of intense study.

Conjecturally, the function space of the latter is known in terms of iterated integrals, or symbols. Using bootstrap ideas, perturbative results at six and seven particles have been obtained at high loop order~\cite{Dixon:2011pw,Dixon:2011nj,Dixon:2013eka,Dixon:2015iva,Dixon:2016nkn,Drummond:2018caf}. This led in particular to insight into how the Steinmann relations are realized in perturbative QFT~\cite{Caron-Huot:2016owq}, and to intriguing observations about a possible cluster algebra structure of the amplitudes~\cite{Golden:2013xva}. 

On the other hand, few results are available to date beyond the planar limit. The four-particle amplitude is known to three loops~\cite{Henn:2016jdu}, and no results are available beyond one loop for more than four particles. In order to study whether properties such as integrability, hidden dual conformal symmetry, and properties of the function space generalize to the full theory, it is crucial to have more data. In this letter, we newly compute, in terms of symbol, a full five-particle scattering amplitude in QFT. While all the required planar master integrals are already known analytically in the literature, one non-planar integral family was still missing, up to now. We fill this gap, and discuss its calculation in a dedicated parallel paper~\cite{paper1}.

\section{Calculation of the master integrals} 

\begin{figure}[t]
  \begin{center}
    \includegraphics[width=0.31\columnwidth]{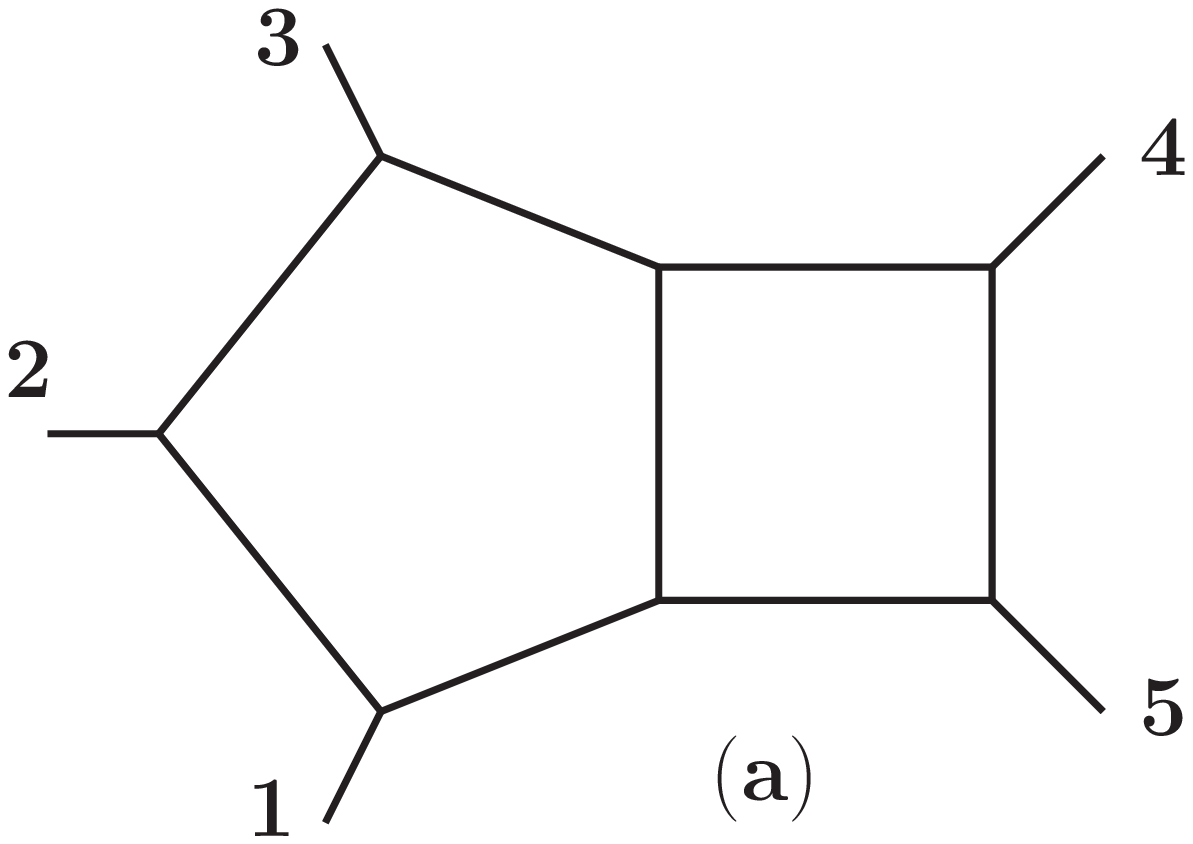}
    \includegraphics[width=0.31\columnwidth]{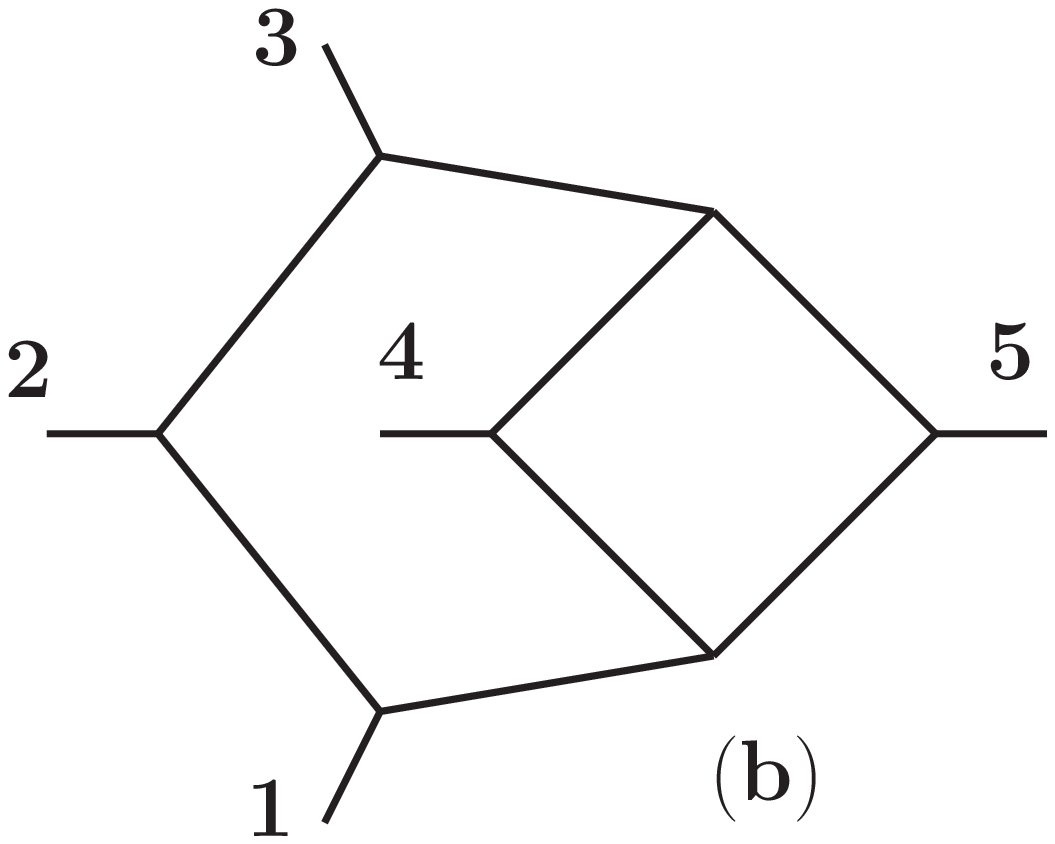}
\includegraphics[width=0.34\columnwidth]{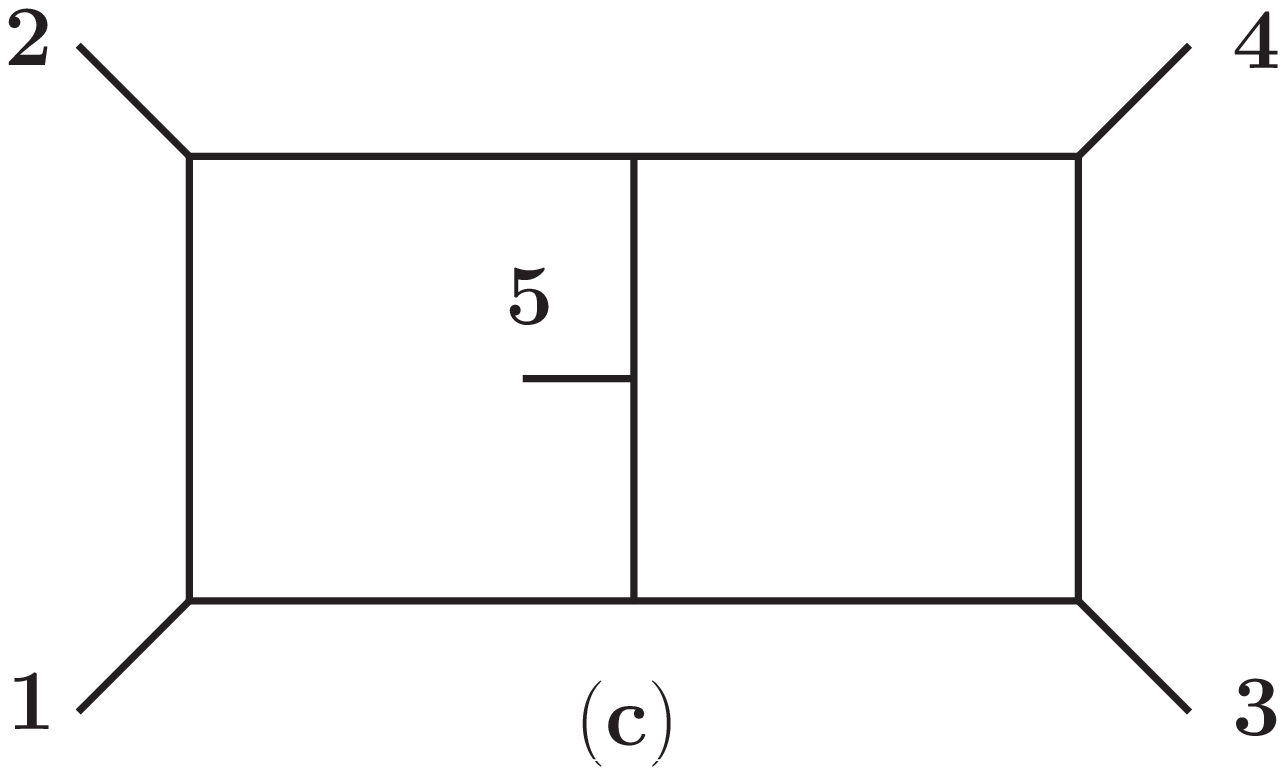}
\includegraphics[width=0.31\columnwidth]{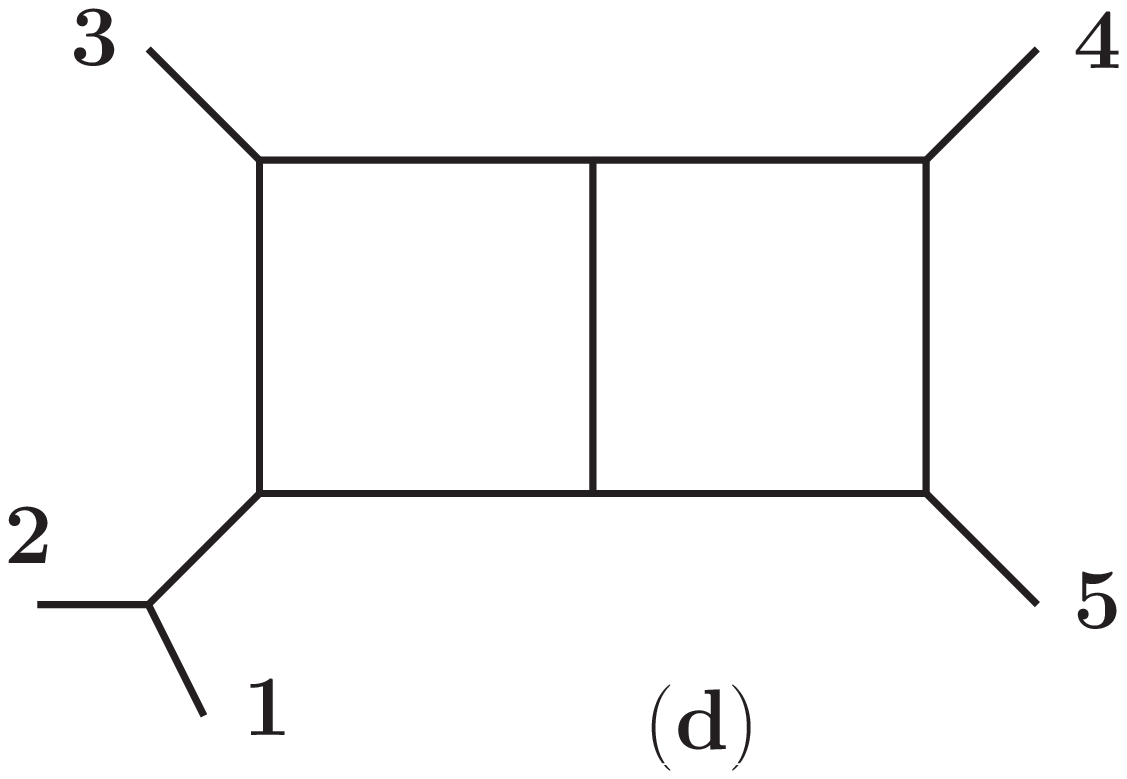}
\includegraphics[width=0.31\columnwidth]{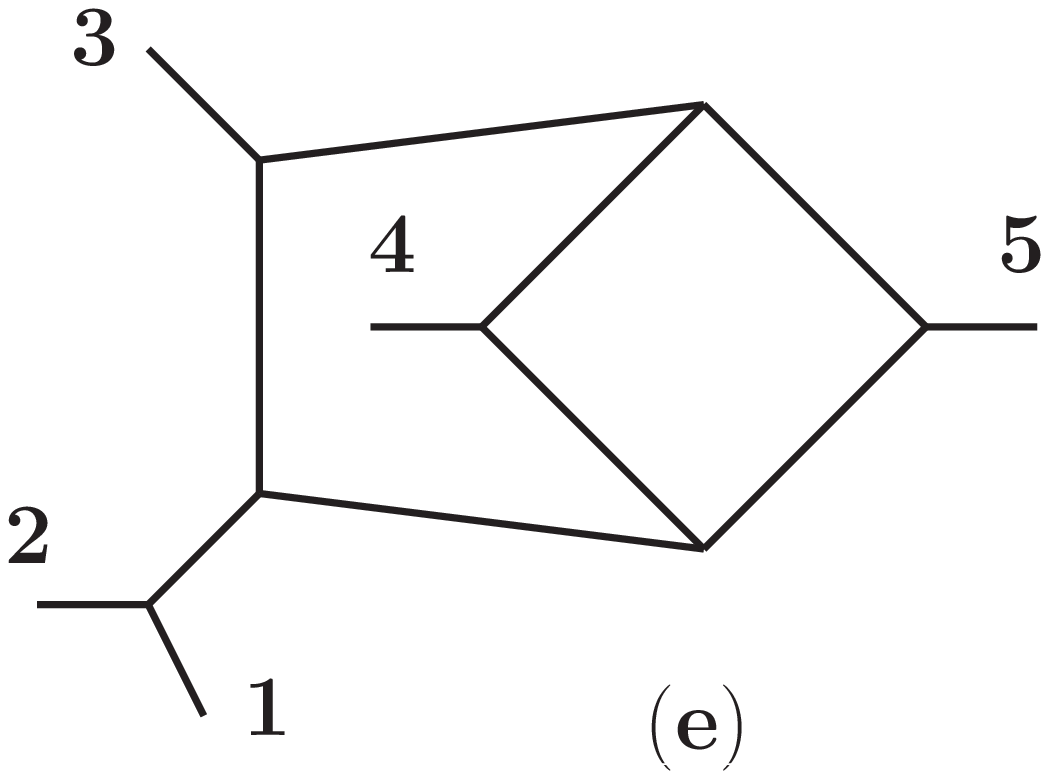}
\includegraphics[width=0.31\columnwidth]{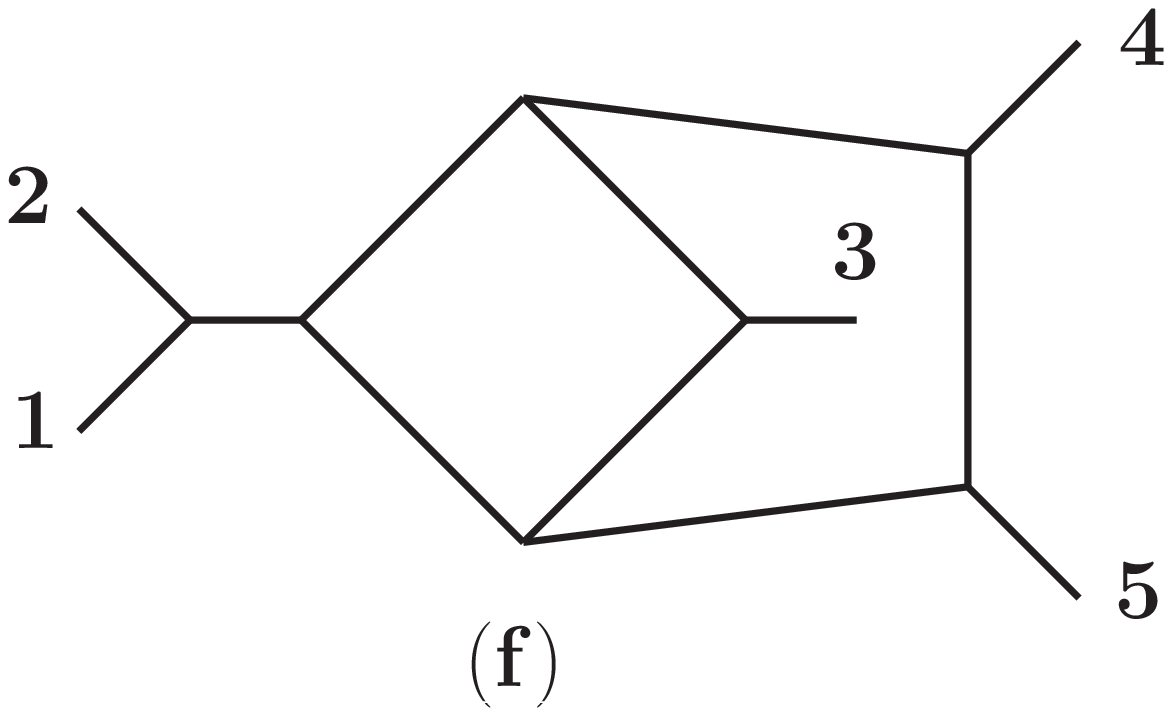}
    \caption{Diagrams in the representation of~\cite{Carrasco:2011mn} of the integrand of the two-loop five-point amplitude in $\mathcal{N}=4$ sYM. We omit the associated numerators and color factors.}
    \label{fig:masterintegrals}
  \end{center}
\end{figure}
The integral topologies needed for massless five-particle scattering at two loops are shown in Fig.~\ref{fig:masterintegrals}.
The integrals in four-point kinematics, Fig.~\ref{fig:masterintegrals}~(d)-(f), are known from refs. \cite{Gehrmann:2000zt,Gehrmann:2001ck}.
The master integrals of the planar topology depicted in Fig.~\ref{fig:masterintegrals}~(a) were computed in ref.~\cite{Gehrmann:2015bfy,Papadopoulos:2015jft,Gehrmann:2018yef}, whereas the non-planar integral family shown in Fig.~\ref{fig:masterintegrals}~(b) was computed in ref.~\cite{Chicherin:2018mue}. (See also \cite{Chicherin:2017dob,Chicherin:2018ubl,Chicherin:2018wes,Abreu:2018rcw}). We devote a parallel paper~\cite{paper1} to the calculation of the missing non-planar family, depicted in Fig.~\ref{fig:masterintegrals}~(c), which we will refer to as double-pentagon. Here we will content ourselves with the details that are directly relevant for the computation of the symbol of the $\mathcal{N}=4$ sYM amplitude.

Genuine five-point functions depend on five independent Mandelstam invariants, $s_{12}$, $s_{23}$, $s_{34}$, $s_{45}$, $s_{51}$, with $s_{ij}=2 p_{i} \cdot p_{j}$.
We will also find the parity-odd invariant $\epsilon_{5} = {\rm tr}[ \gamma_{5} \slashed{p}_{4}\slashed{p}_{5}\slashed{p}_{1}\slashed{p}_{2}]$ useful. Its square can be expressed in terms of the $s_{ij}$ through $\Delta = (\epsilon_{5})^2$, with the Gram determinant 
$\Delta = | 2 p_{i}\cdot p_{j} |$, with $1 \le i,j \le 4$.

The integrals of the double-pentagon topology can be related through Integration-by-Parts relations to a basis of 108 master integrals, which were calculated using the differential equations method~\cite{Gehrmann:1999as,Henn:2013pwa}. In doing this, it was crucial to identify a good basis \cite{Henn:2013pwa,WasserMSc}, namely a basis of integrals with \textit{uniform transcendental weight} (UT integrals): taking into account a conventional overall normalization (extracting a factor $\exp{(-\gamma_{\text{E}}\epsilon)} g^2/(4\pi)^{2-\ep}$ per loop), the order-$1/\eps^4$ terms of such integrals are constant, the order-$1/\eps^3$ terms are given by one-fold integrals (logarithms), and in general the order-$\eps^{-4+n}$ terms are given by $n$-fold iterated integrals.

With this choice of basis, the differential equations assume their canonical form~\cite{Henn:2013pwa}
\begin{align}
\label{canonicalDEpentagon}
d\vec{I}(s_{ij};\epsilon) = \epsilon \left( \sum_{k=1}^{31} a_k d\log W_k(s_{ij}) \right) \vec{I}(s_{ij};\epsilon) \, ,
\end{align}
where $a_k$ are  $108 \times 108$ rational-number matrices, and $W_k$ are the so-called \textit{symbol letters}, algebraic functions of the kinematics encoding the branch-cut structure of the master integrals. 
The emerging \textit{symbol alphabet} coincides with the 31-letter alphabet conjectured in ref.~\cite{Chicherin:2017dob}, and obtained by closing under all permutations of the external momenta the 26-letter alphabet relevant for the planar master integrals~\cite{Gehrmann:2015bfy}.

The master integrals of this canonical basis are thus given by the so-called \textit{pentagon functions}, i.e. iterated integrals in the $31$-letter alphabet of~\cite{Chicherin:2017dob}. 

The construction of the canonical basis was achieved by combining three cutting-edge strategies. The algorithmic search for $\dlog$ integrands, having rational-number leading singularities~\cite{WasserMSc,Chicherin:2018mue}, was in fact supplied, for the highest sector, with two novel methods: a $D$-dimensional analysis of Gram determinants, and the \textit{module lift} computation in algebraic geometry. A thorough discussion is contained in~\cite{paper1}.

Once the differential equations~\eqref{canonicalDEpentagon} and the value of $\vec{I}$ at some boundary point are known, the problem of evaluating the master integrals $\vec{I}$ at any kinematic point in a Laurent expansion around $\eps=0$ is solved~\cite{Henn:2013pwa}. The boundary values can be determined analytically from physical consistency conditions, as discussed in~\cite{Chicherin:2018mue}. In particular, if one is only interested in the \textit{symbol}~\cite{Goncharov:2010jf} of the master integrals $\vec{I}$, the boundary values are needed only at the leading order in the $\epsilon$ expansion, i.e. only at order $1/\epsilon^{2 \ell}$ for a $\ell$-loop integral.
Obtaining the beyond the symbol terms requires applying the method of solving the differential equations of \cite{Chicherin:2018mue,paper1} for all permutations of the integrals appearing in the amplitude, which is beyond the scope of the present paper.
As was already observed for the other two top topologies, the symbols of the master integrals of the double-pentagon satisfy the \textit{second entry condition} conjectured in ref.~\cite{Chicherin:2017dob}.

\section{Calculation of the amplitude}

The integrand for the full five-point two-loop amplitude in $\mathcal{N}=4$ sYM was constructed in~\cite{Carrasco:2011mn} using color-kinematics duality and $D$-dimensional generalized unitarity cuts. In terms of the diagrams shown in Fig.~\ref{fig:masterintegrals}, its expression is very compact
\begin{align}
\label{eq:A_CJ}
\mathcal{A}_5^{(2)} = \sum_{S_5}\left(\frac{I^{(a)}}{2}+\frac{I^{(b)}}{4}+\frac{I^{(c)}}{4}+\frac{I^{(d)}}{2}+\frac{I^{(e)}}{4}+\frac{I^{(f)}}{4} \right),
\end{align}
where the sum runs over all permutations of the external legs.
This representation of the integrand is valid in $D=4-2\epsilon$ dimensions, in the regularization scheme where external states and momenta live in $D=4$ dimensions, and the internal momenta are $D$-dimensional. 

We reduce the diagrams in eq.~\eqref{eq:A_CJ} to the basis of UT integrals for the three top topologies shown in the first row of Fig.~\ref{fig:masterintegrals}. The basis integrals are then substituted with the corresponding symbols, and the permutations are carried out at the symbol level. 

Note that, while having the advantage of being valid in $D$ dimensions, the diagrams figuring in eq.~\eqref{eq:A_CJ} do not have uniform transcendental weight. This complexity in the intermediate stages contrasts with an expected simplicity in the final structure: MHV amplitudes are in fact conjectured to have uniform transcendental weight~\cite{Bern:2005iz,Dixon:2011pw,ArkaniHamed:2012nw,KOTIKOV2007217}, and it is known~\cite{Arkani-Hamed:2014bca} that their leading singularities~\cite{Cachazo:2008vp} are given by Parke-Taylor tree-level super-amplitudes~\cite{Parke:1986,Nair:1988} only,
\begin{align}
\text{PT}(i_1i_2i_3i_4i_5) = \frac{\delta^8(Q)}{\langle i_1 i_2 \rangle \langle i_2 i_3 \rangle \langle i_3 i_4 \rangle \langle i_4 i_5 \rangle \langle i_5 i_1 \rangle  }\,,
\end{align}
where $\delta^8(Q)$ is the super-momentum conservation delta function.
Ref.~\cite{Bern:2015ple} provides a representation of the \textit{four}-dimensional integrand where this property is manifest.

Furthermore, the diagrams in~\eqref{eq:A_CJ} are expressed in terms of MHV prefactors called $\gamma_{ij}$ in~\cite{Carrasco:2011mn}, rather than PT factors. The individual $\gamma_{ij}$, however, can not be uniquely rewritten in terms of PT factors, thus making such structure even more obscure.

In order to suppress the proliferation of spurious rational functions, and to overcome the difficulty in translating the individual $\gamma_{ij}$ MHV prefactors to PT factors, we exploit the insight we have in the structure of the final function, and adopt the following approach. 

While performing the permutations and the sum in eq.~\eqref{eq:A_CJ}, we substitute the kinematic variables with random numbers in the rational prefactors. Then, we single out the prefactor of each individual symbol in the amplitude, and match it with an ansatz made of a $\mathbb{Q}$-linear combination of six independent PT factors.
Following~\cite{Bern:2015ple}, we use a basis of the following six Parke-Taylor factors
\begin{align} \label{Parke-Taylor-basis}
& {\rm PT}_1 = {\rm PT}(12345)\,,\; {\rm PT}_2 = {\rm PT}(12354) \,,\notag \\ 
& {\rm PT}_3 = {\rm PT}(12453) \,,\; {\rm PT}_4 = {\rm PT}(12534) \,, \\ 
& {\rm PT}_5 = {\rm PT}(13425) \,,\; {\rm PT}_6 = {\rm PT}(15423)\,. \notag
\end{align}

Finally, the coefficients of the ans\"atze for the rational prefactors of the individual symbols appearing in the amplitude are fixed entirely by considering six random sets of kinematics. Additional sets are used to validate the answer.

After summing over all permutations, therefore, the underlying simplicity of the full amplitude emerges: all spurious rational functions cancel out, and the amplitude turns out to be a linear combination of UT integrals, with prefactors given by PT tree-level super-amplitudes.

The amplitude is a vector in color space. The color structures of the diagrams in eq.~\p{eq:A_CJ} are obtained by associating a structure constant $i \sqrt{2} f^{abc}$ with each trivalent vertex in Fig.~\ref{fig:masterintegrals}. 
We prefer to expand the amplitude in a basis $\{ \cT_{\lambda} \}$ of 12 single-traces, $\lambda=1,\ldots,12$, and 10 double-traces, $\lambda=13,\ldots,22$, defined in eqs.~(2.1) and (2.2) of~\cite{Edison:2011ta}. E.g. 
\begin{align}
& \cT_1 = \text{Tr}(12345)-\text{Tr}(15432)\, ,\nt 
& \cT_{13} = \text{Tr}(12)\left(\text{Tr}(345)-\text{Tr}(543)\right)\, , \label{tdef}
\end{align} 
where $\text{Tr}(i_1 i_2...i_n)$ denotes the trace of the generators $T^a$ of the fundamental representation of $SU(N_c)$ normalized as $\text{Tr}(T^a T^b) = \delta^{ab}$. The other color basis elements $\cT_{\lambda}$ are given by permutations of $\cT_1$ and $\cT_{13}$.

Adopting the conventions of ref.~\cite{Edison:2011ta}, we decompose the amplitude as follows 
\begin{align}
\mathcal{A}_5^{(2)} = \sum_{\lambda=1}^{12} \left(N_c^{2} A_{\lambda}^{(2,0)} +A_{\lambda}^{(2,2)} \right) \cT_{\lambda}+\sum_{\lambda=13}^{22} \left(N_c A_{\lambda}^{(2,1)} \right) \cT_{\lambda}\,.
\end{align}
All partial amplitudes $A_{\lambda}^{(2,k)}$ exhibit the elegant structure discussed above
\begin{align}
A_{\lambda}^{(2,k)} = \frac{1}{\epsilon^4} \sum_{w=0}^{4} \epsilon^w \sum_{i=1}^6  \text{PT}_i \, f_{w,i}^{(k,\lambda)} + {\cal O}(\eps)\,,
\end{align}
where $\text{PT}_i$ are the PT factors defined by eqs.~\eqref{Parke-Taylor-basis}, $f_{w,i}^{(k,\lambda)}$ are weight-$w$ symbols.

Our result was validated through a series of strong checks, that we describe below.

\subsection{Color relations}
The partial amplitudes $A_{\la}^{(2,k)}$ satisfy group-theoretic relations, which automatically follow from rearranging the color structure of the amplitude in the basis~$\{ \cT_\la \}$. As a result, the most color-subleading part of the two-loop amplitude $A_\la^{(2,2)}$ can be rewritten as a linear combination of the planar $A_\la^{(2,0)}$ and of the double-trace $A_\la^{(2,1)}$ components~\cite{Edison:2011ta}.

\subsection{ABDK/BDS ansatz}	
We verified that the leading-color partial amplitudes $A_{\lambda}^{(2,0)}$, $\lambda=1,\ldots,12$, match the formula proposed in refs.~\cite{Anastasiou:2003kj,Bern:2005iz}, and can thus be
obtained by exponentiating the one-loop amplitude~\cite{Bern:1993mq}. The ABDK/BDS ansatz was previously confirmed numerically~\cite{Bern:2006vw,Cachazo:2006tj}, and was shown to follow from a dual conformal Ward identity~\cite{Drummond:2007au}. 

\subsection{Collinear limit}
We consider the limit in which the momenta of two particles, say 4 and 5, become collinear, i.e. we let $p_4 = z P$ and $p_5 = (1-z) P$, with $P = p_4+p_5$. In this limit the two-loop five-point amplitude factorizes into a universal color-blind splitting amplitude and a 4-point amplitude~\cite{Bern:2004cz}. Choosing particles 4 and 5 to be positive helicity gluons, we have
\begin{align}
&
\left(\cA^{(2)}_5\right)^{a_1,a_2,a_3,a_4,a_5} \overset{4||5}{\to} f^{a_4 a_5 b} \Bigl[ {\rm Split}^{(0)}_-(z;4^+,5^+)\, \cA_4^{(2)} \nt
&\hspace{1.5cm} + N_c \,{\rm Split}^{(1)}_-(z;4^+,5^+)\, \cA_4^{(1)} \nt 
&\hspace{1.5cm} + N^2_c \,{\rm Split}^{(2)}_-(z;4^+,5^+)\, \cA_4^{(0)} \Bigr]^{a_1,a_2,a_3,b} \,,\label{collim}
\end{align} 
where ${\rm Split}^{(\ell)}_-(z;4^+,5^+)$ and $\mathcal{A}^{(\ell)}_4$ are the $\ell$-loop splitting amplitude and 4-point amplitude $123P$ respectively.
In order to control the collinear limit $4||5$,
we introduce a parameter $\delta$ which approaches 0 in the limit, and $y$, which stays finite, and use the following momentum twistor-inspired parametrization for the Mandelstam invariants
\begin{align}
&s_{12} = \frac{s x \sqrt{y}}{x \sqrt{y}+\delta(1+x)+\delta^2 \sqrt{y}(1+x)} \nt
&s_{23} = s x \nt 
&s_{34} = \frac{s z}{1+(1+x)\sqrt{y}(1-z)\delta}, \nt
& s_{45} = \frac{s x (1+x) \sqrt{y}\delta^2}{x \sqrt{y} + \delta(1+x) + \delta^2 \sqrt{y}(1+x)} \nt
& s_{15} = \frac{s x (1-z)}{1+(1+x)(1-z)\sqrt{y}\delta} \label{params}
\end{align}
where  $s,t$ are Mandelstam invariants of the four-point amplitude $123P$, and $x = t/s$. Substituting the parametrization~\p{params} into the letters of the pentagon alphabet, and expanding them up to the leading order in $\delta$, yields a 14-letter alphabet. Note however that the right-hand side of eq.~\p{collim} contains only the letters $\{\delta, s, x,1+x,z,1-z\}$. The symbol of the four-point amplitude in fact belongs to the alphabet $\{x,1+x\}$, and the loop corrections of the splitting factors are specified by the alphabet $\{z,1-z\}$. This means that the majority of the 14-letter alphabet has to drop out in the collinear limit, thus making this cross-check very constraining.
We used the two-loop splitting amplitudes given in~\cite{Bern:2004cz}, and the four-point amplitude up to $\mathcal{O}(\eps^2)$ from~\cite{Henn:2016jdu}, and found perfect agreement with eq.~\eqref{collim}.

\subsection{Infrared dipole formula and hard remainder function}
Up to two loops, the IR singularities of gauge-theory scattering amplitudes of massless particles factorize according to the \textit{dipole formula}~\cite{Catani:1998bh,Aybat:2006wq,Aybat:2006mz,Almelid:2015jia} 
\begin{align} \label{renorm}
\cA(s_{ij},\ep) = \mathbf{Z}(s_{ij},\ep) \cA^{f}(s_{ij},\ep)\,,
\end{align} 
where the factor $\mathbf{Z}(s_{ij},\ep)$ captures all IR singularities, and $\cA^{f}$ is thus a finite hard part of the five-point amplitude $\cA \equiv \cA_5$. 
We use bold letters to indicate operators in color space. Since we are interested in the symbol of the amplitude we omit all beyond-the-symbol terms in the following formulae. 
The factor $\mathbf{Z}(s_{ij},\ep)$ is then given by
\begin{align} 
\mathbf{Z}(s_{ij},\ep) = \exp g^2 \left( \frac{\mathbf{D}_0}{2\ep^2} - \frac{\mathbf{D}}{2\ep} \right) \,,
\end{align} 
where $\mu$ is a factorization scale, and the dipole operators acting on pairs of incoming particles are defined by
\begin{align}
\mathbf{D}_0 = \sum_{i\neq j} \vec{\mathbf{T}}_i \cdot \vec{\mathbf{T}}_j \,, \;\;
\mathbf{D} = \sum_{i\neq j} \vec{\mathbf{T}}_i \cdot \vec{\mathbf{T}}_j \, \log\left( -\frac{ s_{ij}}{\mu^2} \right),
\end{align}
with $\mathbf{T}_i^b \circ T^{a_i} = - i f^{b a_i c_i} T^{c_i}$.

Let us denote by $\cA^{(\ell)}_{;w}$ the weight-$w$ part of the $\ell$-loop amplitude, which is of order $\eps^{w-2\ell}$ in the $\eps$-expansion of $\cA^{(\ell)}$. Then, we find that the IR-divergent terms of $\cA^{(2)}$ are completely determined by the lower-loop data as dictated by the dipole formula~\p{renorm}
\begin{align}
&\cA^{(2)}_{;0} = \frac{25}{2} N_c^2 \, \cA^{(0)}  \,, \;\; \cA^{(2)}_{;1} = \frac{5}{2} N_c\, \mathbf{D} \, \cA^{(0)}
\,, \notag \\ 
& \cA^{(2)}_{;2} = \frac{1}{8} \left[\mathbf{D} \right]^2 \, \cA^{(0)} + 5 N_c \, \cA^{(1)}_{;2} \,, \notag \\
& \cA^{(2)}_{;3} = \frac{1}{2} \mathbf{D} \, \cA^{(1)}_{;2} + 5 N_c \, \cA^{(1)}_{;3} \,,
\end{align}
and the two-loop correction ${\cal H}^{(2)}$ to the IR-safe hard function ${\cal H}(s_{ij}) \equiv \lim\limits_{\eps \to 0} A^{f}(s_{ij},\ep)$
is given by 
\begin{align} 
\cA^{(2)}_{;4} = {\cal H}^{(2)} + 5 N_c \, \cA_{;4}^{(1)} + \frac{1}{2} \mathbf{D} \, \cA^{(1)}_{;3}\,.
\end{align}
We note that the symbol of ${\cal H}^{(2)}$ does not depend on $W_{31}$.
\newline

The two-loop double-trace part of the hard function ${\cal H}(s_{ij})$ is the truly new piece of information.
The IR poles and the leading-color components of the amplitude are in fact entirely determined by lower loop information through the dipole formula~\eqref{renorm} and the ABDK/BDS ansatz~\cite{Anastasiou:2003kj,Bern:2005iz} respectively. Moreover, the most-subleading-color part can be obtained from the leading-color and the double-trace components via color relations~\cite{Edison:2011ta}. Only the double-trace part of the hard function can be considered as new, and
it is therefore worth looking for a more compact representation of it. 

We find the following concise formula
\begin{align}
\label{eq:Hrep2}
\mathcal{H}^{(2)}_{\text{dbl-tr}} = \sum_{S_5}\Bigl[N_c \, \cT_{13}  \text{PT}_1\, g^{(4)}_{\text{seed}} \Bigr]\,,
\end{align}
where $g_{\text{seed}}^{(4)}$ is a weight-4 symbol, $\text{PT}_1$ is defined by eq.~\eqref{Parke-Taylor-basis}, and $\cT_{13}$ is defined in eq.~\p{tdef}. We provide the expression of $g^{(4)}_{\text{seed}}$ split into parity-even and odd part in the ancillary files $\mathtt{Hdt\_seed\_even.txt}$ and $\mathtt{Hdt\_seed\_odd.txt}$, respectively.

\section{Multi-Regge limit}
We now study the multi-Regge limit~\cite{Kuraev:1976ge,DelDuca:1995hf} of the amplitude in the physical $s_{12}$-channel
\begin{align}
s_{12} \gg s_{34} > s_{45} > 0 \, ,  \qquad  s_{23} < s_{15} < 0\,.
\end{align}
We parametrize the kinematics in this limit as
\begin{align}
& s_{12} = s/x^2 \,,\;\;
s_{34} = s_1/x \,,\;\; 
s_{45} = s_2/x \,,\nt
& \hspace{1.5cm} s_{23} = t_1 \,,
\;\; 
s_{15} = t_2 \,,
\end{align}
and let $x \to 0$.
Substituting this parametrization in the pentagon alphabet, and expanding up to the leading order in $x \to 0$, we find that it reduces significantly, and factorizes into the tensor product of four independent alphabets:  $\{ x \}$, $\{ \kappa \}$, $\{ s_1 , s_2, s_1 - s_2 , s_1 + s_2 \}$, $\{ z_1, z_2 , 1 - z_1 , 1- z_2 , z_1 - z_2, 1-z_1 - z_2 \}$,
where $\kappa$, $z_1$ and $z_2$ are defined as
\begin{align}
\kappa = \frac{s_1 s_2 }{s} \,, \,  t_1 = - \kappa z_1 z_2 \,,\ t_2 = -\kappa (1-z_1)(1-z_2) \,.
\end{align}
The two one-letter alphabets simply correspond to powers of logarithms. The third alphabet corresponds to harmonic polylogarithms~\cite{Remiddi:1999ew}, and the fourth to two-dimensional harmonic polylogarithms~\cite{Gehrmann:2001jv}.

The Regge limit of the single-trace leading-color terms has already been studied~\cite{Bartels:2008ce}. The simple form of the ABDK/BDS formula~\cite{Anastasiou:2003kj,Bern:2005iz} at five points, consisting only of logarithms, is in fact Regge-exact.

We are now for the first time in the position to take the multi-Regge limit of the double-trace subleading-color part of the hard function ${\cal H}^{(2)}_{\text{ dbl-tr}}$, and we find that it vanishes at the symbol level. It will be interesting to investigate whether this remains true at function level. 

We can also go further, and consider the subleading power corrections to ${\cal H}^{(2)}_{\text{dbl-tr}}$, of which we present analytically the leading-logarithmic contribution
\begin{align*}
\mathcal{H}^{(2)}_{\text{dbl-tr}} \underset{x\to 0}{\longrightarrow} \; &\frac{2}{3} x \log^4 (x) \biggl[ \frac{\kappa z_2}{s_1}\bigl( 11 (\cT_{15} + \cT_{19})  - 4 \cT_{14} \bigr) \notag\\ 
&+ \frac{\kappa (1- z_1)}{s_2}\bigl( 11 ( \cT_{16} +\cT_{21}) - 4 \cT_{17} \bigr) \biggr]\,.
\end{align*}
We provide the weight-4 symbol of the first subleading power corrections to ${\cal H}^{(2)}_{\text{dbl-tr}}$ in the ancillary file $\mathtt{subleading\_
multi\_Regge.txt}$.

\section{Conclusions and outlook}
In this letter, we computed for the first time the symbol of a two-loop five-particle amplitude analytically. The infrared divergent part of our result constitutes a highly non-trivial check of the two-loop dipole formula for infrared divergences, leading to the first analytic check of two-loop infrared factorization for five particles. Our result provides a substantial amount of analytical data for future studies. For example, we started the analysis of the multi-Regge limit at subleading color. We found that the leading power terms vanish, and provided the subleading terms. Further terms can be straightforwardly obtained from our symbol. We observed that the non-planar pentagon alphabet implies a simple structure of the Regge limit. It will be interesting to understand whether this alphabet is also sufficient to describe five-particle scattering at higher loop orders. It will also be relevant to explore whether hints of directional dual conformal symmetry~\cite{Chicherin:2018wes,Bern:2017gdk,Bern:2018oao}, which is present at the level of individual integrals, can be found at the level of the full amplitude, and whether there is a connection to Wilson loops~\cite{Ben-Israel:2018ckc}.
\\

{\it Note added:}
While this manuscript was in the final stage of preparation, the preprint~\cite{Abreu:2018aqd} appeared. The authors of~\cite{Abreu:2018aqd} use another set of master integrals to calculate the symbol of the two-loop five-point amplitude in $\mathcal{N}=4$ sYM, in agreement with our result. \\

\section{Acknowledgments}
We thank V.~Mitev for collaboration in early stages of this work.
This research received funding from Swiss National Science Foundation (Ambizione grant PZ00P2 161341), the European Research Council (ERC) under the European Union's
Horizon 2020 research and innovation programme (grant agreement No 725110), {\it Novel
structures in scattering amplitudes}. J. H., Y. Z. and S. Z. also wish to thank the Galileo Galilei Institute for hospitality during the workshop ``Amplitudes in the LHC era".

\bibliographystyle{h-physrev} 

\bibliography{5point_refs}

\end{document}